\documentclass[journal]{IEEEtran}
  \usepackage{amssymb}
  \usepackage{latexsym}
  \usepackage{amsfonts}
  \usepackage{amsthm}
  \usepackage{amsmath}
  \usepackage{graphics}
  \usepackage{setspace}
  \usepackage{graphicx}
  \usepackage{epstopdf}
  \usepackage[usenames,dvipsnames,svgnames,table]{xcolor}
  \usepackage{float}
  \usepackage{wrapfig}
  \usepackage{color}
  \usepackage{rotating}
  \usepackage{array} 
  \usepackage[hyphens]{url}
  \usepackage[table]{xcolor}
  \usepackage{multirow}  
  \usepackage{adjustbox}

  \usepackage{longtable}

  \usepackage{forloop,supertabular}
\newcounter{count}
\newcommand{\comment}[1]{}

\definecolor{SW}{RGB}{87,157,28}
\definecolor{SH}{RGB}{153,51,102}
\definecolor{SC}{RGB}{197,0,11}    
\definecolor{SE}{RGB}{0,132,209}
\definecolor{SN}{RGB}{255,149,14}

\newcommand{\catDxxWearable}{\adjustbox{valign=m}{\colorbox{SW}{}}}
\newcommand{\catExxSmartHome}{\adjustbox{valign=m}{\colorbox{SH}{}}}
\newcommand{\catAxxSmartCity}{\adjustbox{valign=m}{\colorbox{SC}{}}}
\newcommand{\catCxxEnvironment}{\adjustbox{valign=m}{\colorbox{SN}{}}}
\newcommand{\catBxxEnterprise}{\adjustbox{valign=m}{\colorbox{SE}{}}}

\begin{document}
%
\title{The Emerging Internet of Things Marketplace From an Industrial Perspective: A Survey}
%
%
%

\author{Charith~Perera,~\IEEEmembership{~Member,~IEEE,}     
        Chi Harold Liu~\IEEEmembership{Member,~IEEE},
        Srimal~Jayawardena,~\IEEEmembership{~Member,~IEEE,}   

\thanks{This work is sponsored in part by National Natural Science Foundation of China (Grant No.: 61300179).}
\thanks{Charith~Perera, and Srimal~Jayawardena,  are with the Research School of Computer Science, The Australian National University, Canberra, ACT 0200, Australia. (e-mail: firstname.lastname@ieee.org)}
\thanks{Chi Harold  Liu is with Beijing Institute of Technology, China. (e-mail: chiliu@bit.edu.cn)}
\thanks{Manuscript received xxx xx, xxxx; revised xxx xx, xxxx.}}

%
%


\markboth{IEEE Transactions on Emerging Topics in Computing}%
{Shell \MakeLowercase{\textit{et al.}}: Bare Demo of IEEEtran.cls for Journals}
%



\maketitle

\begin{abstract}
The Internet of Things (IoT) is a dynamic global information network consisting of internet-connected objects, such as Radio-frequency identification (RFIDs), sensors, actuators, as well as other instruments and smart appliances that are becoming an integral component of the future internet. Over the last decade, we have seen a large number of the IoT solutions developed by start-ups, small and medium enterprises, large corporations, academic research institutes (such as universities), and private and public research organisations making their way into the market. In this paper, we survey over one hundred IoT smart solutions in the marketplace and examine them closely in order to identify the technologies used, functionalities, and applications. More importantly, we identify the trends, opportunities and open challenges in the industry-based the IoT solutions.  Based on the application domain, we classify and discuss these solutions under five different categories: smart wearable, smart home, smart, city, smart environment, and smart enterprise. This survey is intended to serve as a guideline and conceptual framework for future research in the IoT and to motivate and inspire further developments. It also provides a systematic exploration of existing research and suggests a number of potentially significant research directions.
\end{abstract}




\section{Introduction}
\label{sec:Introduction}

The Internet of Things (IoT) is a network of networks where, typically, a massive number of objects/things/sensors/devices are connected through communications and information infrastructure to provide value-added services.  The term was first coined in 1998  and later defined as ``The Internet of Things allows people and things to be connected Anytime, Anyplace, with Anything and Anyone, ideally using Any path/ network and Any service'' \cite{ZMP007}.  As highlighted in the  definition, connectivity among the devices is a critical functionality that is required to fulfil the vision of the IoT. The main reasons behind such interest are the capabilities and sophistication that the IoT will bring to society \cite{P003}. It promises to create a world where all the objects around us are connected to the Internet and communicate with each other with minimal human intervention. The ultimate goal is to create ``a better world for human beings'', where objects around us know what we like, what we want, and what we need, and hence act accordingly without explicit instructions \cite{P040}.

There have been a number of surveys conducted in the IoT domain. The area of the IoT has been broadly surveyed by Atzori et al. in \cite{P003}. Bandyopadhyay et al. have surveys of the IoT middleware solutions in \cite{P118}. Layered architecture in industrial IoT are discussed in \cite{TII01}. A similar survey focusing on data mining techniques for the IoT are discussed in \cite{Z1039}. Edge mining in IoT paradigm is discussed in \cite{TII10}. In contrast to the traditional data mining, edge mining takes place on the wireless, battery-powered, and smart sensing devices that sit at the edge points of the IoT. The challenges in self organizing in IoT are discussed in \cite{TII02}. Atzori et al \cite{TII07} have discussed how smart objects can be transformed in to social objects. Such transformation will allow the network to enhance the level of trust between objects that are `friends' with each other. IoT technologies and solutions towards Smart Cities are reviewed in \cite{TII04}. Communication protocols and technologies play a significant role in IoT. Sheng et al. \cite{TII06} have survey a protocol stack developed specifically for IoT domain by Internet Engineering Task Force (IETF).

 Internet of things: vision, applications and research challenges are discussed from a research perspective in \cite{Z1037, Z1038}. Further, the IoT has been surveyed in a context-aware perspective by Perera et al. \cite{ZMP008}. A survey on facilitating  experimentally IoT research is presented by \cite{Z1040}. Palattella et al. \cite{Z1041} have introduced a communications protocol stack to support and standardise IoT communication. Security challenges such as general system security, network security, and application security in the IoT are discussed in \cite{Z1042}. The security issues in perception layer, network layer and application layer in architectures have discussed in \cite{TII03}. Hardware devices, specially nano sensors and technologies, used in IoT are surveyed in \cite{TII09}. Another similar survey has been done by Hodges et al. \cite{Z1055}. This paper discusses a open-source hardware platform called {.NET Gadgeteer}, a rapid prototyping platform for small electronic gadgets and embedded hardware devices.{.NET Gadgeteer} is coming from an industrial setting similar to Arduino \cite{P411}.

Besides the above articles, there are a number of surveys and reviews that have been conducted by researchers around the world in the IoT domain, from which we have hand picked some to represent the existing body of knowledge.

As far as we know, however, no survey has focused on IoT industry solutions. All the above-mentioned surveys have reviewed the solutions proposed by the academic and research community and refer to scholarly publications. In the present paper, we review the IoT solutions that have been proposed, designed, developed, and brought to market by industrial organisations. These organisations range from start-ups and small and medium enterprises to large corporations. Because of their industrial and market-driven nature, most of the IoT solutions in the market are not published as academic works. Therefore, we collected information about the solutions from their respective web-sites, demo videos, technical specifications, and consumer reviews. Understanding how technologies are used in the IoT solutions in the industry's marketplace is vital for academics, researchers, and industrialists so they can identify trends, opportunities, industry requirements, demands, and open research challenges. It is also critical for understanding trends and open research gaps so future research directions can be guided by them.


The present paper is organised into sections as follows:  In Section \ref{sec:Evaluation}, we evaluate and examine the functionalities provided by each solution under the five categories identified in the earlier section. At the end of that section, we summarise the functionalities and highlight the major  domains that are commonly targeted by the solutions. Then, we examine the IoT solutions from a technology and business perspective. Hardware platforms, software platforms, additional equipment, communication protocols, and the energy sources used by each solution are examined in Section \ref{sec:Technology_Review}. At the end of that section, we summarise the technologies and business models used by the IoT solutions so trends and opportunities can be identified. In Section \ref{sec:Lessons_Learned}, we identify such trends using the evaluations we conducted in the previous sections. Later, opportunities for research and development will be assessed in Section \ref{sec:Ope_Research_Challenges}. Concluding remarks
will be presented in Section \ref{sec:Conclusions}.

\section{Functionality Review of IoT Solutions}
\label{sec:Evaluation}

In this section, we focus on the functionalities of the IoT solutions. The next section discusses the technologies used by these solutions under common themes. In both sections, our intention is not to describe each IoT solution in detail, but to organise them into common themes so we can identify trends and opportunities. However, readers can use citation numbers to track a given IoT solution throughout the paper, if desired. Such an option allows consolidating the knowledge we have put separately in two sections, to better understand a single IoT solution. In Section \ref{sec:Lessons_Learned}, we will analyse the trends from both the functional and the technological point of views.

\subsection{Smart Wearable}
\label{sec:SS:Smart_Wearable}

\begin{figure}[t!]
 \centering
 \includegraphics[scale=1.2]{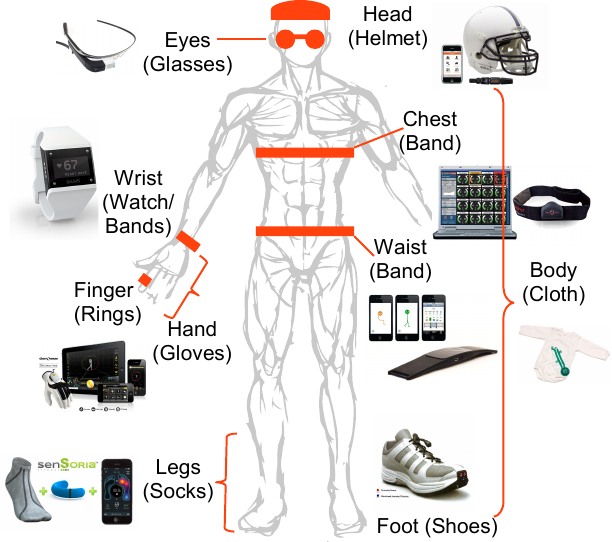}
 \caption{Different body parts popularly targeted by wearable IoT solutions in the industry market-place.}
 \label{Figure:Bodyparts}	
\vspace{-0.3cm}	
\end{figure}

Wearable solutions are diverse in terms of functionality. They are designed for a variety of purposes as well as for wear on a variety of parts of the body, such as the head, eyes, wrist, waist, hands, fingers, legs, or embedded into different elements of attire.  In Table \ref{Tbl:Wearable_IoT_Solutions}, we summarise popular wearable IoT solutions. This table includes a brief description of each solution, context information gathered, similar solutions, and the context-aware functionality provided by the solution. The IoT solutions are categorised by the body part on which the solution must be worn, as illustrated in Figure \ref{Figure:Bodyparts}. In addition to the industry IoT solutions, academic solutions in the wearable computing area are discussed in \cite{Z1048, Z1049}. Challenges and opportunities in developing smart wearable solutions are presented in \cite{Z1046}.

\begin{table*}[htbp]
\centering
\footnotesize
\caption{Summary of Wearable IoT Solutions} 
\label{Tbl:Wearable_IoT_Solutions}
\begin{tabular}{  c p{15.4cm} }
\hline
 & \begin{center}
 Functionalities Provided by Different wearable IoT solutions
 \end{center}  \\ \hline \hline

\multirow{3}{*}{ \begin{sideways}Cloth \hspace{1.0cm}     \end{sideways}}  & 
 
 \begin{itemize}
 \item Monitor respiration, body position, activity level, skin temperature, and audio of a baby using pressure, stretch, noise, and temperature sensors, and provide notification through a smart phone regarding any situation that parents need to attend to (Baby Monitor: RestDevice / Mimobaby  \cite{RestDevice}).
 
\item A sleep-tracking device that uses a thin-film sensor strip placed on a mattress in combination with smart phone to help to create a nightly rest profile. It helps to improve user's sleep over time (Sleep Tracking: Beddit \cite{Beddit}).
 
  \item  Jacket relieves anxiety and stress from those diagnosed with autism spectrum disorder (ASD) or attention-deficit/hyperactivity disorder. Built-in motion sensors and pressure sensors track the frustration and activity levels of the child throughout the day and generate custom notification alerts based on that information (Medical Assistant :MyTJacket \cite{TJacket}).
 \end{itemize}  \\

 \multirow{3}{*}{ \begin{sideways}Waist / Chest \hspace{0.3cm}     \end{sideways}}  & 
  
  \begin{itemize}
\item  Tracks posture and daily activities in real time. It provides advice on posture issues so users can improve their posture (Daily Activity and Fitness Monitor / Medical: Lumoback \cite{LUMOback}). 

\item A device that updates Twitter when a baby in the womb kicks its mother (Medical Assistant: kickbee \cite{CoreyMenscher}). 

\item A chest band that tracks heart rate, speed, distance, stress level, calories, and activity level. It allows recommended working out within certain heart rate zones to achieve goals such as weight loss or cardiovascular improvement. (Personal Sports Assistant: BioHarness \cite{BioHarness}).
  \end{itemize}  \\

 \multirow{3}{*}{ \begin{sideways}Wrist \hspace{1.4cm}     \end{sideways}}  & 
  
  \begin{itemize}
 \item A wrist band that tracks steps taken, stairs climbed, calories burned, and hours slept, distance travelled, and quality of sleep and provides recommendation for a healthier lifestyle (Daily Activity and Fitness Monitor: MyBasis \cite{BASIS}, BodyMedia \cite{LINKArmband}, Lark \cite{lark}). 
 
  \item Open wearable sensor platform, a wrist band that comprises number of different sensors such as pulse, blood flow sounds, blood oxygen saturation, blood flow waveform, pulse, acceleration, type of activity, calories burned and number of steps taken, skin temperature (Open Platform: AngelSensor  \cite{angelsensor}). 
  
   \item EMBRACE+, a wrist band that connects to the user's smartphone via Bluetooth and displays any notifications user may receive as ambient light notifications (Personal Sports Assistant: EmbracePlus \cite{embraceplus}). 
   
    \item Electrocardiogram technology (ECG), Bluetooth connectivity and a suite of sensors are used to recognize users' heart rhythm uniquely and securely and continuously log into users' nearby devices (Secure Authentication: nymi \cite{getnymi}). 
    
     \item A watch that helps athletes to keep track of their training. Context information such as mapping, distance, speed, heart rate, and light are collected and fused to generate athletes' training profile (Personal Sports Assistant: Leikr \cite{leikr}).
  \end{itemize}  \\ 
  
   \multirow{3}{*}{ \begin{sideways}Eyes \hspace{1.4cm}     \end{sideways}}  & 
    
    \begin{itemize}
\item  Sports-specific (skiing) goggles that monitor jump analytics, speed, navigation, trip recording, and peer tracking (Personal Sports Assistant: Oakley Goggles \cite{OakleyAirwaveGoggles}). 

\item A pair of glasses that consist of camera, projector, and sensors to support functionalities such as navigation calendar notification, navigation, voice activated, voice translation, communication and so on. It also acts as an open platform where different context-ware functionalities can be built using provided sensors and processing capabilities (Open Platform: Google Glass \cite{glass}).
    \end{itemize}  \\ 
    
     \multirow{3}{*}{ \begin{sideways}Head \hspace{1.4cm}     \end{sideways}}  & 
      
      \begin{itemize}
\item  Sports-specific (American football) helmet that determines when to take a player off the field and seek medical advice through impact detection and analysis (Personal Sports Assistant: TheShockBox \cite{shockboxImpactalertsensors}). 

\item A bicycle helmet that detects a crash. If the user's head hits the pavement (or anything hard (ice, snow, dirt)), a signal will be sent to the smartphone automatically to generate a call for help (Emergency Accident monitor: ICEdot \cite{icedot}).
      \end{itemize}  \\ 
      
       \multirow{3}{*}{ \begin{sideways}Hands \hspace{1.4cm}     \end{sideways}}  & 
        
        \begin{itemize}
\item Monitor, analyze and improve golf swing through motion sensors embedded in gloves (Personal Sports Assistant: Zepp \cite{Zepp} ) 

\item A ring that monitors and keeps track of the user's heart rate (Medical Assistant: ElectricFoxy \cite{electricfoxy}).
        \end{itemize}  \\ 
        
         \multirow{3}{*}{ \begin{sideways}Legs / Foot \hspace{1.4cm}     \end{sideways}}  & 
          
          \begin{itemize}
\item  A sock that combines an accelerometer with textile sensors to measure steps, altitude and calories burnt. It helps runners to avoid potentially dangerous techniques: heel striking or excessive forefoot running that could lead to back pain or Achilles ten-don injuries. (Daily Activity and Fitness Monitor / Medical: Heapsylon \cite{SensoriaSmartSock}) 

\item A pair of shoes that provides feedback through vibrations in an intuitive and non-obstructive way. The shoes suggest the right direction and detect obstacles (Disability Assistance: LeChal \cite{LeChal})
          \end{itemize}  \\ 
          
           \multirow{3}{*}{ \begin{sideways}Internal \hspace{1.4cm}     \end{sideways}}  & 
            
            \begin{itemize}
\item A small patch worn on the body working together with 1mm sensor-enabled pills and a back-end cloud service to collect and process real-time information (e.g. heart rate, temperature, activity and rest patterns throughout the day) on the user's medication adherence (Medical: Proteus Digital Health \cite{proteusdigitalhealth}).
            \end{itemize}  \\ 
            
             \multirow{3}{*}{ \begin{sideways}Multi \hspace{1.4cm}     \end{sideways}}  & 
              
              \begin{itemize}
\item A device that can be worn on multiple body parts tracks steps taken, stairs climbed, calories burned, and hours slept, distance travelled, quality of sleep (Daily Activity and Fitness Monitor: Fitbit \cite{fitbit}). 

\item An ultra-small GPS unit and five in-built sensors are used to collect data and fused to tell the camera exactly the right moment to take photos (Leisure: Autographer \cite{Autographer}). 

\item Remote monitoring system that collects data through devices that can be worn on different body parts on a patient's physiological conditions to support physicians (Health Monitoring: Preventice BodyGuardian \cite{BodyGuardianRemoteMonitoringSystem}).
              \end{itemize}  \\

 \hline
\end{tabular}
\label{Tbl:Smart_Wearables_Summary}
\end{table*}

\subsection{Smart Home}
\label{sec:SS:Smart_Home}
Solutions in this category make the experience of living at home more convenient and pleasant for the occupants. Some smart home \cite{Z1053} solutions also focus on assisting elderly people in their daily activities and on health care monitoring \cite{Z1050}. Due to the large market potential, more and more smart home solutions are making their way into the market. From the academic point of view, smart energy and resource management \cite{Z1051, Z1061}, human--system interaction \cite{Z1052}, and activity management \cite{Z1054}, have been some of the major foci. 

\textit{\textbf{Platforms: }}Smartthings \cite{SmartThingsHomeWatchSolution} is a generic platform that consists of hardware devices, sensors, and software applications. Context information is collected through sensors and injected into applications where reasoning and action are performed accordingly. For example, the sprinkler installed in the user's garden can detect rain and turn itself off to save energy. Ninjablocks \cite{ninjablocks} and Twine \cite{Twine} provide similar functionalities. These solutions were mainly developed to support smart home and building domains, but they can be customised to other domains. HomeOS \cite{Z1058} is a platform that supports home automation. Instead of custom hardware (e.g. a smartthings hub), HomeOS is a software platform which can be installed on a normal PC. As with the smartthings platform, applications can be installed to support different context-aware functionalities (e.g. capturing an image from a door camera and sending it to the user when someone rings the doorbell). Lab-of-things \cite{Z1059} is a platform built for experimental research. It allows the user to easily connect hardware sensors to the software platform and enables the collection of data and the sharing of data, codes, and participants. 

\textit{\textbf{Virtual Assistance: }}Ubi \cite{Ubi} supports residents by acting as a voice-activated computer. It can perform tasks such as audio calendar, feed reader, podcast, voice memos, make lighting-based notifications to indicate the occurrence of certain events, weather, stock, email, and so on. Ubi has a microphone and speakers. It also has sensors to monitor the environment, such as monitoring the temperature, humidity, air pressure, and ambient light. Netatmo \cite{Netatmo} is an air quality monitoring solution for smart homes. In order to determine air quality, it collects context information from sensors such as temperature, humidity, and CO2. The solution monitors the home environment and sends an alert when the residents' attention is required. Meethue \cite{meethue} is a bulb which can be controlled from mobile devices. The bulb reacts to the context and can change its colour and brightness according to user preferences, time / day / season, and activity (e.g. resident enters home) and is also sensitive to changes in the weather during the day. 

\textit{\textbf{Smart Objects:} }— WeMo \cite{WeMoSwitc} is a Wi-Fi enabled switch that can be used to turn electronic devices on or off from anywhere. Context-aware schedules are also supported, where turning on or off is performed automatically according to the time of day, sunrise, or sunset. Tado \cite{TADO} is an intelligent heating control that uses a smartphone. It offers context-aware functionalities such as turning down the heating when the last person leaves the house, turning the heating back up before someone gets home, and heats the house less when the sun is shining. Nest \cite{NestThermostat} is a thermostat that learns what temperatures users like and builds a context-aware personalised schedule. The thermostat automatically turns to an energy-efficient `away temperature' when occupants leave the home. If it senses activity, such as a friend's coming over to water the plants, Nest could start warming up the house. The thermostat can be activated remotely through the Nest mobile app. Lockitron \cite{Lockitron} is a door lock that can be opened and closed by a phone over the Internet. Residents can authorise family and friends to open a given door by providing authorisation over the Internet, so that others can use their smartphones to unlock doors. Blufitbottle \cite{BlueFit} is a water bottle that records drinking habits while keeping the users healthy and hydrated. If the user starts to fall behind with hydration, the bottle has customisable sounds and lights to alert them. 

\textit{\textbf{Digital Relationships:}} — Wheredial \cite{WhereDial} offers a way to make a personal connection with family members or friends. It retrieves a person's location from Foursquare, Google Latitude, and a variety of other services. Then it rotates the dial (like a clock) to show where the person is at a given moment. Goodnightlamp \cite{GoodNightLamp} is a family of connected lamps that let the user remotely communicate the act of coming back home to their loved ones easily and in an ambient way by fusing location-aware sensing. The objective of Wheredial and Goodnightlamp is the same: helping to build and maintain family relationships and further strengthen friendships by mitigating the fact that the users are apart from each other. Such solutions are extremely important  in terms of  social, psychological, and mental well-being.

\subsection{Smart City}
\label{sec:SS:Smart_City}

Towns and cities accommodate one-half of the world's population, creating tremendous pressure on every aspect of urban living. Cities have large concentrations of resources and facilities \cite{P535}. The enormous pressure towards efficient city management has triggered various Smart City initiatives by both government and private sector businesses to invest in information and communication technologies to find sustainable solutions to the growing problems \cite{ZMP008}. Smart grid is one of the domains in which academia, industry, and governments are interested and invested significantly \cite{Z1064, Z1067}.

\textit{\textbf{Smart Traffic}}  ParkSight \cite{ParkSight} is a parking management technology designed for cities. Context information is retrieved through sensors (magnetometers) embedded in parking slots. Application support is provided via location and map services to guide drivers to convenient parking based on real-time context analysis. Uber \cite{Uber} allows users to request a ride at any time. The company in a particular place sends a cab. In contrast to transitional taxi services, no phone call or pick-up location is required. A mobile application shows the cabs close to the users and their movement in real time. A cab can be requested by means of a single smartphone tap. Alltrafficsolutions \cite{AllTraffic} collects traffic data through sensors and visualises it on maps in order to provide drivers with traffic updates. Further, it provides remote equipment management support related to traffic control (e.g. changes in digital road signs, speed limit boards, variable message signs (e.g. `event parking') to drivers, and changes in the brightness of digital signs based on the context information). Streetbump \cite{StreetBum} is a crowd-sourcing project that helps residents to improve their neighbourhood streets. Volunteers use the Streetbump mobile application to collect road condition data while they drive. The data are visualised on a map to alert residents regarding real-time road conditions. The collected data provide governments with real-time information with which to fix problems and plan long-term investments.

\textit{\textbf{Platforms}} Libelium \cite{WaspMote} provides a platform of low-level sensors that is capable of collecting a large amount of context information to support different application domains [9]. Thingworx \cite{thingworx} and Xively \cite{xively} are cloud-based on-line platforms that process, analyse, and manage sensor data retrieved through a variety of different protocols.

\textit{\textbf{Resource Management}} — SmartBelly \cite{SmartBellycomponents} is a smart waste management solution. It provides a sensor-embedded trash can that is capable of real-time context analysis and alerting the authorities when it is full and needs to be emptied. Location information is used to plan efficient garbage collection. Echelon \cite{SmartStreetLighting} has developed a smart street lighting solution transforming street-lights into intelligent, energy-efficient, remotely managed networks. It schedules lights to be turned on or off and sets the dimming levels of individual lights or groups of lights so a city can intelligently provide the right level of lighting needed by analysing the context such as time of day, season, or weather conditions.

\textit{\textbf{Activity Monitoring}}  Livehoods \cite{Livehoods} offers a new way to conceptualise the dynamics, structure, and character of a city by analysing the social media its residents generate. This is achieved through collecting context information such as check-in patterns. Livehoods shows how citizens use the urban landscape and other resources. Scenetap \cite{SceneTap} shows real-time info about the city's best places. It shows the context information of a given location such as how many people are there, the male to female ratio, and the average age of everyone inside. This helps users to find the best places to hang out (e.g. cinema, bar, restaurant) at a given time and gives information such as availability.

\subsection{Smart Environment}
\label{sec:SS:Smart_Environment}

\textit{\textbf{Air Quality Monitoring}} — Airqualityegg \cite{AirQualityEgg} is a community-led sensor system that allows anyone to collect context information such as the carbon monoxide (CO) and nitrogen dioxide (NO2) gas concentrations outside their home. Such data are related to urban air pollution. Communitysensing \cite{CommonSense} is also an air quality monitoring system which provides both hand-held devices and a platform to be fixed into municipal vehicles such as street sweepers. Aircasting \cite{AirCasting} is a platform for recording, mapping, and sharing health and environmental data using smartphones and custom monitoring devices. Context information includes sound levels, temperature, humidity, carbon monoxide (CO) and nitrogen dioxide (NO2) gas concentrations, heart and breathing rate, activity level, and peak acceleration.

\textit{\textbf{Water Quality Monitoring}} — Floating Sensor Network \cite{TheFloatingSensorNetwork} collects real-time, high-resolution data on waterways via a series of mobile sensing `drifters' that are placed in the water. It collects context information such as water quality, water flow movement, and speed, temperature and water pollution.  Intelligentriver \cite{IntelligentRiver} is also an observation system that supports research and provides real-time monitoring, analysis and management of water resources. A similar solution has been developed by Roboshoal \cite{Shoal}. The difference is that their station is a mobile fish-shaped robotic device whose movement is controllable. Dontflush \cite{dontflushme} is designed to enable residents to understand when overflows happen and reduce their waste-water production before and during an overflow event. Context information is processed in order to determine real-time sewage levels and advise users regarding safe flushing through a context-aware light bulb and SMS.


\textit{\textbf{Natural Disaster Monitoring}} — AmritaWNA \cite{AmritaWNA} is a wireless landslide detection system that is capable of releasing alerts about possible landslides caused by torrential rain in the region. Context information is collected by sensors such as strain gauge piezometers, vibrating wire piezometers, dielectric moisture sensors, tilt meters, and geophones. This is a station-based solution. Insightrobotics \cite{ComputerVisionWildfireDetectionSystem} is a solution that detects forest fires by fusing context information collected through various kinds of sensors (i.e. temperature, wind, and so on) and networked cameras.

\textit{\textbf{Smart Farming}} — Microstrain \cite{ShelburneVineyardRemoteMonitoring} has developed a wireless environmental sensing system to monitor key conditions during the growing season in vineyards. Context information such as current temperature and soil moisture conditions, leaf wetness, and solar radiation is collected and fused in order to monitor vineyards remotely and alert farmers regarding critical situations. The collected data are used to support both real-time context-aware functionalities and historic data analysis. Bumblebee \cite{Bumblebeenestingproject} monitors the lives of bumblebees by collecting and processing context information such as visual, audio, temperature, sunlight, and weather. It automatically tweets the current situation of the colony and well-being of the bees. Hydropoint \cite{HydroPoint} retrieves context information through 40,000 weather stations and automatically schedules irrigation based on individual landscape needs and local weather conditions, resulting in lower water bills and energy savings.

\begin{table*}[t]
\centering
\footnotesize
\renewcommand{\arraystretch}{1.05}

\caption{Summary of the Taxonomy used in Table~\ref{Tbl:Evaluation_of_Previous_Research_Efforts}} 
\label{Tbl:Evaluation_of_Technology_Taxonomy}
\vspace{-0.3cm}
\begin{tabular}{ c l m{13cm} }
\hline
 & Taxonomy & Description \\ \hline \hline
%
1 & Project & The name of the project, product or solution  sorted by `Category' and then by `Project Name' within each category in ascending order \\ 
3 & Year &  Last known active year of the project. \\ 
4 & Category & Category that the solution belongs to. 
Each category is denoted by a different colour: 
\iftrue
red {\color{SC}\rule{0.2cm}{0.2cm}}   (smart city), yellow 
{\color{SN}\rule{0.2cm}{0.2cm}}  (smart environment), blue {\color{SE}\rule{0.2cm}{0.2cm}}  (smart enterprise), green  {\color{SW}\rule{0.2cm}{0.2cm}} (smart wearable), and purple {\color{SH}\rule{0.2cm}{0.2cm}} (smart home).
\else
red \catC  (smart city), yellow \catV   (smart environment), blue \catN  (smart enterprise), green \catW (smart wearable), and purple \catH (smart home).
\fi
Some solutions belongs to multiple categories.\\ 
5 & Availability & The ability to obtain (free or purchase)  each IoT solution. Available solutions are denoted by ($\checkmark$). . \\ 
6 & Price & The price of the IoT solution. It can be unit price or service prise or both. All the prices are denoted in US dollars. Superscript number denotes the original currency where the IoT solution has been sold: USD (1),  AUD (2), EURO (3), and GBP (4). Further, if it is a service, additional superscripts are used to denote payment period: monthly (m) and yearly (y). Currency conversion has been performed on 2013-December-11 using online service provided by www.xe.com. \\
7 & Hardware and / or Software & This indicate whether the IoT solution consists of hardware (H) unit, software (S) service or both \\ 
8 & Wireless Technology & Different types of wireless technologies used in each of the IoT solutions are denoted as follows: (Mobile) Ad-hoc Network using Ultrasonic communication (A) , WiFi (W), Bluetooth (B), USB (U), Celluar Radio / GSM (C), ZigBee  (Z), RF  (R), GPS (G)\\
9 & Platform & IoT solutions have utilized different platforms. Some solutions support multiple platforms as follows: Android (A),    Blackberry  (K), IOS (I), Web based service (B), Mac OSX (M), Windows (W), and Linux   (L)\\ 
10 & License & The IoT solutions are covered by different licenses as follows: Commercial  (C), Open-source  (O), Research \& Development (R), and Free    (F). 
No specific license information were available for cases denoted by (R)   which were carried out as research initiatives and possibly available for collaborative research work with permission for non-commercial work. 
(F) denotes solutions which are available for free without any governing licenses.\\ 
11 & Unit and / or subscription  & The IoT solution that are sold as a unit are denoted by (U) and others cases where the solution need to be purchased as a subscription are denoted by (S) \\ 
12 & Product and / or Service & IoT solutions that marketed as a product  are denoted by (P) and the solutions marketed as services are denoted by (S). Some IoT solutions have both product and service components.  \\ \hline
\multicolumn{3}{c}{Note: Cases where sufficient information were not available are denoted by (-)}
\end{tabular}
\label{Tbl:Summarized taxonmy}
\vspace{-16pt}
\end{table*}

\subsection{Smart Enterprise}
\label{sec:SS:Smart_Enterprise}

In general, enterprise IoT solutions are designed to support infrastructure and more general purpose functionalities in industrial places, such as management and connectivity.

\textit{\textbf{Transportation and Logistics}} — Senseaware \cite{SenseAware} is a solution developed to support real-time shipment tracking. The context information such as location, temperature, light, relative humidity and biometric pressure is collected and processed in order to enhance the visibility of the supply chain. HiKoB \cite{PROJECTGRIZZLY} collects real-time measurements such as temperature gradients within the road, current outdoor temperatures, moisture, dew and frost points from sensors deployed in roads and provides traffic management, real-time information on traffic conditions, and services for freight and logistics. Cantaloupesys \cite{SeedPlatform} allows the user to keep track of stocks in vending machines remotely. Timely and optimal replenishment strategies (i.e. the elimination of unnecessary truck travel and smaller loads per truck) are determined from context information related to usage patterns.

\textit{\textbf{Infrastructure and Safety}}— SmartStructures \cite{SmartPile} collects data from sensors embedded within concrete piles in foundations which enables post-construction long-term load and event monitoring. Yanzi \cite{RemoteSiteManagement} is a solution that enables the user to monitor, maintain, and manage lifts, elevators, heating systems, energy consumption, motion detection, and surveillance. Context information is retrieved through sensors such as video, temperature, motion, and light.  Engaugeinc \cite{RemoteFireExtinguisherMonitoringSystem} is a remote fire extinguisher monitoring system. Multiple sensors are used to collect context information that allows the user to determine when a fire extinguisher is blocked, when it is missing from its designated location, or when its pressure falls below safe operating levels. Alerts are sent out via email, phone, pager, and a software-based control panel.

\textit{\textbf{Energy and Production}} — Wattics \cite{Smartmetering} is a smart metering solution that manages energy consumption at the individual appliance and machine level. Context information is used to understand usage pattern recognitions of each appliance through software algorithms which predict and load balance to reduce the energy cost. Sightmachine \cite{SightMachine} continuously processes context data gathered from sensors, lasers, and network cameras, makes assessments in real time, and allows the user to stop problems before they happen with regard to industrial manufacturing machines and equipment. 

\textit{
\textbf{Resources Management}}— Onfarmsystems \cite{OnFarm} is an IoT solution designed to facilitate smart farming through accommodating increasingly complex and interconnected farming equipment. Context information such as energy, pesticide, mapping/ location, soil moisture, telemetry, weather, and monitoring are used to support efficient real-time decision-making. HeatWatch \cite{HeatWatchII} is a cattle monitoring solution that records the activities of each animal. Recorded context information includes such information as movement, time of day of the mount, and duration of the mount. Such information enables farmers to breed more cows and heifers earlier, obtain better results (more pregnancies), use less semen, spend much less time, and be more efficient. Motionloft \cite{Motionloftpropertyanalytics} is a solution that monitors pedestrian and vehicle movements in real-time by collecting activity data. It enables boutique retailers, large chains, restaurants, and bars to understand the impact which vehicle and pedestrian traffic has on their revenue.

\section{Technology Review of IoT Solutions}
\label{sec:Technology_Review}

In this section, we summarise, from the technology point of view, the results surveyed  so far. Our review criteria are explained in detail in Table \ref{Tbl:Evaluation_of_Technology_Taxonomy}. The results are presented in Table \ref{Tbl:Evaluation_of_Previous_Research_Efforts}. Specifically, our objective is not to discuss the technologies in details but to survey and compare the usage of the technologies (i.e. column (7), (8), (9) in Table \ref{Tbl:Evaluation_of_Technology_Taxonomy}) and solution models employed by different IoT solutions (i.e. column (10), (11), (12) in Table \ref{Tbl:Evaluation_of_Technology_Taxonomy}). Over the last few years, IoT has been surveys in many different perspective. Leading surveys that discuss different aspects of IoT technologies are presented in Section \ref{sec:Introduction}. In Section \ref{sec:Lessons_Learned}, we will analyse trends and lessons learned from the survey in detail.

\begin{table*}[t!]
\caption{Evaluation of Surveyed Research Prototypes, Systems, and Approaches}
\footnotesize
\renewcommand{\arraystretch}{0.8}
\begin{tabular}{
 m{4.5cm} 
 c 
 m{0.4cm}  
 c
 p{0.2cm}
 c
 c
 c
 c
 c
 c
 c
 } 
 \hline
 Project Name     &          
 \begin{sideways}Citations \end{sideways}   &   
 Year &                                         
 \begin{sideways}Category\end{sideways} &                                         
 \begin{sideways}Availability \end{sideways} &   
  Price &        
 \begin{sideways}\begin{minipage}[b]{1.5cm}Hardware  Software \end{minipage} \end{sideways} & 
 \begin{sideways}\begin{minipage}[b]{1.6cm}Wireless \\Technology \end{minipage} \end{sideways} &  
 \begin{sideways}Platform \end{sideways} &            
 \begin{sideways}License \end{sideways} &        
 \begin{sideways}\begin{minipage}[b]{1.6cm}Unit  \\Subscription \end{minipage} \end{sideways} &     
 \begin{sideways}\begin{minipage}[b]{1.6cm}Product  \\Service \end{minipage} \end{sideways}    
 \\  
 \hline \hline 
(1)      & (2)              & (3)      & (4)      &(5)            & (6)       & (7)      & (8)    & (9)                            &  (10)   &  (11)   & (12)                \\


.NET Gadgeteer   & \cite{Z1055} & 2013 & \catAxxSmartCity \catBxxEnterprise \catCxxEnvironment \catExxSmartHome \catDxxWearable & $\checkmark$ & Vary & H   & All  & -  & O  & U & P \\
Arduino  & \cite{P411} & 2013 & \catAxxSmartCity \catBxxEnterprise \catCxxEnvironment \catExxSmartHome \catDxxWearable & $\checkmark$ & Vary & H   & All  & -  & O  & U & P \\
ThingWorx  & \cite{thingworx} & 2013 & \catAxxSmartCity \catBxxEnterprise \catCxxEnvironment  & $\checkmark$ & - & S   & All  & B  & C  & S & S \\
Xively  & \cite{xively} & 2013 & \catAxxSmartCity \catBxxEnterprise \catCxxEnvironment   & $\checkmark$ & 999 - 39,000$^{y}$ & S   & All  & B  & C  & S & S \\

All Traffic & \cite{AllTraffic} & 2013 & \catAxxSmartCity & $\checkmark$ & - & H,S & B,C,W & A,B & C & S & P,S \\ 
CityDashboard & \cite{CityDashboard} & 2013 & \catAxxSmartCity & $\checkmark$ & Free & S & C,W & B & R & - & S \\ 
Common Sense & \cite{CommonSense} & 2013 & \catAxxSmartCity & $\checkmark$ &  & H,S & B,C & B & R & U & P,S \\ 
Enevo & \cite{Enevo} & 2013 & \catAxxSmartCity & $\checkmark$ & - & H & C & B & - & S & P,S \\ 
Estimote Beacons & \cite{EstimoteBeacons} & 2013 & \catAxxSmartCity & $\checkmark$ & 99.00$^{1}$ & H,S & B & A,I & O & U & P \\ 
Livehoods & \cite{Livehoods} & 2013 & \catAxxSmartCity & $\checkmark$ & Free & S & C,W & B & F & - & S \\ 
ParkSight & \cite{ParkSight} & 2013 & \catAxxSmartCity & - & - & - & - & - & - & - & - \\ 
Pavegen & \cite{Pavegen} & 2013 & \catAxxSmartCity & $\checkmark$ & - & H & - & B & R & U & P \\ 
Placemeter & \cite{Placemeter} & 2013 & \catAxxSmartCity & - & - & H,S & C,W & - & - & S & S \\ 
Points  & \cite{Points} & 2013 & \catAxxSmartCity & $\checkmark$ & - & H & W & B & - & U & S \\ 
SceneTap & \cite{SceneTap} & 2013 & \catAxxSmartCity & $\checkmark$ & - & H,S & C,W & A,B,I & C & S & S \\ 
Smart Street Lighting & \cite{SmartStreetLighting} & 2013 & \catAxxSmartCity & $\checkmark$ & - & - & - & - & - & - & - \\ 
SmartBelly components & \cite{SmartBellycomponents} & 2013 & \catAxxSmartCity & $\checkmark$ & - & - & C & B & - & - & P,S \\ 
Street Bum & \cite{StreetBum} & 2013 & \catAxxSmartCity & $\checkmark$ & Free & S & C,W & I & F & - & S \\ 
Uber & \cite{Uber} & 2013 & \catAxxSmartCity & $\checkmark$ & - & - & C & A,I & C & U & S \\ 
WaspMote & \cite{WaspMote} & 2013 & \catAxxSmartCity \catBxxEnterprise & $\checkmark$ & - & H,S & C,R,W,Z & B & - & - & - \\ 
Motionloft property analytics & \cite{Motionloftpropertyanalytics} & 2013 & \catAxxSmartCity\catBxxEnterprise & $\checkmark$ & 279.00$^{1,m}$ & H,S & - & B & C & S & S \\ 
PROJECT GRIZZLY & \cite{PROJECTGRIZZLY} & 2013 & \catAxxSmartCity\catBxxEnterprise & $\checkmark$ & - & H,S & - & - & O & - & - \\ 
Air Quality Egg & \cite{AirQualityEgg} & 2013 & \catAxxSmartCity\catCxxEnvironment & $\checkmark$ & - & H,S & - & B & O & P & S \\ 
AirCasting  & \cite{AirCasting} & 2013 & \catAxxSmartCity\catCxxEnvironment & $\checkmark$ & - & H,S & B,C,W & A & O & U & P,S \\ 
FleetSafer OBD & \cite{FleetSaferOBD} & 2013 & \catBxxEnterprise & $\checkmark$ & - & H & B & - & C & U & P \\ 
GPS Trailer Tracking & \cite{GPSTrailerTracking} & 2013 & \catBxxEnterprise & $\checkmark$ & - & H & G,W & - & C & U & P \\ 
HeatWatch II  & \cite{HeatWatchII} & 2006 & \catBxxEnterprise & $\checkmark$ & - & H,S & R & - & C & U & P,S \\ 
Intelligence Golf Course Irrigation & \cite{IntelligenceGolfCourseIrrigation} & 2012 & \catBxxEnterprise & $\checkmark$ & - & H,S & - & - & - & - & P \\ 
Limitless Wireless Operator  & \cite{LimitlessWirelessOperatorInterface} & \multicolumn{1}{l}{-} & \catBxxEnterprise & $\checkmark$ & - & H & R & - & C & U & P \\ 
OnFarm & \cite{OnFarm} & 2012 & \catBxxEnterprise & $\checkmark$ & 0-1,500$^{1,y}$ & H,S & - & - & C & S & P,S \\ 
Asset Tracking   System & \cite{RecovereAssetTrackingandRecoverySystem} & \multicolumn{1}{l}{-} & \catBxxEnterprise & $\checkmark$ & - & H & C,R & - & C & U & P \\ 
Remote Fire Extinguisher   & \cite{RemoteFireExtinguisherMonitoringSystem} & \multicolumn{1}{l}{-} & \catBxxEnterprise & $\checkmark$ & - & H,S & - & - & C & U & P \\ 
Remote Site Management & \cite{RemoteSiteManagement} & 2011 & \catBxxEnterprise & $\checkmark$ & - & H,S & W & A,B,I & C & S & P,S \\ 
Remote Tank Monitoring Solution & \cite{RemoteTankMonitoringSolution} & 2013 & \catBxxEnterprise & $\checkmark$ & - & H,S & - & - & C & U & P \\ 
Seed Platform & \cite{SeedPlatform} & 2012 & \catBxxEnterprise & $\checkmark$ & - & H,S & - & B & C & S & P,S \\ 
SenseAware & \cite{SenseAware} & 2013 & \catBxxEnterprise & $\checkmark$ & - & H,S & C & B & C & U & P,S \\ 
Sight Machine & \cite{SightMachine} & 2013 & \catBxxEnterprise & $\checkmark$ & - & H,S & - & - & O & - & - \\ 
Smart metering & \cite{Smartmetering} & 2011 & \catBxxEnterprise & $\checkmark$ & - & H,S & - & - & C & U & P,S \\ 
Smart Pallet & \cite{SmartPallet} & 2012 & \catBxxEnterprise & $\checkmark$ & - & H & R & - & C & U & P \\ 
SmartPile & \cite{SmartPile} & 2013 & \catBxxEnterprise & $\checkmark$ & - & H,S & - & B & C & U & P,S \\ 
temperaturealert & \cite{temperaturealert} & 2012 & \catBxxEnterprise & $\checkmark$ & - & H & C,U,W & - & C & U & P \\ 
Bumblebee nesting project & \cite{Bumblebeenestingproject} & 2013 & \catCxxEnvironment & $\checkmark$ & - & - & - & B & R & - & S \\ 
Wildfire Detection System & \cite{ComputerVisionWildfireDetectionSystem} & 2012 & \catCxxEnvironment & $\checkmark$ & - & H,S & - & - & - & U & P,S \\ 
dontflushme & \cite{dontflushme} & 2013 & \catCxxEnvironment & - & - & H,S & - & B & O & U & P,S \\ 
Intelligent River & \cite{IntelligentRiver} & 2013 & \catCxxEnvironment & $\checkmark$ & - & H,S & - & - & R & U & P,S \\ 
Vineyard Remote Monitoring & \cite{ShelburneVineyardRemoteMonitoring} & 2013 & \catCxxEnvironment & $\checkmark$ & - & H,S & C & - & C & - & - \\ 
Shoal & \cite{Shoal} & 2012 & \catCxxEnvironment & $\checkmark$ & 32,879.13$^{4}$ & H,S & A & - & C & U & P,S \\ 
The Floating Sensor Network & \cite{TheFloatingSensorNetwork} & 2013 & \catCxxEnvironment & $\checkmark$ & - & H,S & C & B & R & U & P,S \\ 
Asthmapolis & \cite{Asthmapolis} & 2013 & \catDxxWearable & - & - & H,S & B & I,A & R & U & P \\ 
Autographer & \cite{Autographer} & 2013 & \catDxxWearable & $\checkmark$ & 399.00$^{1}$ & H & B,U & - & C & U & P \\ 
BASIS & \cite{BASIS} & 2012 & \catDxxWearable & $\checkmark$  & 199.00$^{1}$ & H,S & B,W & A,I,B & C & U & P \\ 
BEARTek Gloves & \cite{BEARTekGloves} & 2012 & \catDxxWearable & - & - & H & B & - & R & U & P \\ 
Beddit & \cite{Beddit} & 2013 & \catDxxWearable & $\checkmark$ & 411.26$-680.85^{3}$ & H,S & B & A,I & C & U,S & P \\ 
BleepBleep & \cite{BleepBleep} & 2013 & \catDxxWearable & $\checkmark$ & - & H,S & B & A,I & C & U,S & P,S \\ 
BodyGuardian Remote Monitoring & \cite{BodyGuardianRemoteMonitoringSystem} & 2013 & \catDxxWearable & - & - & H,S & B,W & B & - & - & P,S \\ 
fitbit & \cite{fitbit} & 2013 & \catDxxWearable & $\checkmark$  & 63.94-137.08$^{2}$ & H,S & B,W & A,I,M,W & C & U & P \\ 
Galaxy Gear & \cite{fitbit} & 2013 & \catDxxWearable & $\checkmark$  & 299.00$^{1}$  & H,S& B,W & A & C & U & P \\
Helios Bars & \cite{HeliosBars} & 2013 & \catDxxWearable & $\checkmark$ & 199.00$^{1}$ & H,S & B,G & I & C & U & P \\ 
LINK Armband & \cite{LINKArmband} & 2013 & \catDxxWearable & $\checkmark$ & 149.00$^{1}$ & H,S & B & A,I,B & C & U,S & P,S \\ 
Lively & \cite{Lively} & 2013 & \catDxxWearable & $\checkmark$ & 149.00$^{1}$- & H,S & C & B,I & C & U,S & P,S \\ 
LUMOback & \cite{LUMOback} & 2013 & \catDxxWearable & $\checkmark$  & 149.00$^{1}$ & H,S & B & I & C & U & P \\ 
MUZIK headphones & \cite{MUZIKheadphones} & 2013 & \catDxxWearable & - & 299.00$^{1}$ & H,S & B & A,I & O & U & P \\ 
NFC ring & \cite{NFCring} & 2013 & \catDxxWearable & - & - & H,S & B & - & O & U & P \\ 
Oakley Airwave Goggles & \cite{OakleyAirwaveGoggles} & 2013 & \catDxxWearable & $\checkmark$  & 599.95$^{1}$ & H,S & B & A,I & C & U & P \\ 
Owlet  & \cite{Owlet} & 2013 & \catDxxWearable & - & - & H,S & - & I & C & U & P \\ 
Rest Device & \cite{RestDevice} & 2012 & \catDxxWearable & - & - & H,S & W & - & R & U & P \\ 
Sensoria Smart Sock & \cite{SensoriaSmartSock} & 2013 & \catDxxWearable & - & - & H,S & B & - & R & U & P \\ 
shockbox Impact alert sensors & \cite{shockboxImpactalertsensors} & 2013 & \catDxxWearable & $\checkmark$ & 149.99$^{1}$ & H & B & A,I,K & C & U & P \\ 


\hline
\label{Tbl:Evaluation_of_Previous_Research_Efforts}

\end{tabular}
\end{table*}

\begin{table*}[t!]
\centering
\footnotesize
\renewcommand{\arraystretch}{0.8}
\begin{tabular}{
 m{4.8cm} 
 c 
 m{0.8cm}  
c
 p{0.4cm}
 c
 c
 c
 c
 c
 c
 c
 } 
 \hline
 Project Name     &          
 \begin{sideways}Citations \end{sideways}   &   
 Year &                                         
 \begin{sideways}Category\end{sideways} &                                         
 \begin{sideways}Availability \end{sideways} &   
  Price &        
 \begin{sideways}\begin{minipage}[b]{1.5cm}Hardware  Software \end{minipage} \end{sideways} & 
 \begin{sideways}\begin{minipage}[b]{1.6cm}Wireless \\Technology \end{minipage} \end{sideways} &  
 \begin{sideways}Platform \end{sideways} &            
 \begin{sideways}License \end{sideways} &        
 \begin{sideways}\begin{minipage}[b]{1.6cm}Unit  \\Subscription \end{minipage} \end{sideways} &     
 \begin{sideways}\begin{minipage}[b]{1.6cm}Product  \\Service \end{minipage} \end{sideways}    
 \\  
 \hline \hline 
(1)      & (2)              & (3)      & (4)      &(5)            & (6)       & (7)      & (8)    & (9)                            &  (10)   &  (11)   & (12)                \\



SIGMO & \cite{SIGMO} & 2013 & \catDxxWearable & $\checkmark$ & - & H,S & B & - & C & U & P \\ 
TJacket & \cite{TJacket} & 2013 & \catDxxWearable & $\checkmark$ & 499.00$^{1}$ & H,S & - & I,A & C & U & P \\ 
Withings Wireless Scales & \cite{WithingsWirelessScales} & 2013 & \catDxxWearable & $\checkmark$  & 137.48-206.26$^{3}$ & H,S & B,W & A,I & R & U & P \\ 
BeClose Senior Safety System & \cite{BeCloseSeniorSafetySystem} & 2013 & \catDxxWearable\catExxSmartHome & $\checkmark$  & 399.00$^{1}$- & H,S & W & - & C & U,S & P \\ 
Smart pill bottles & \cite{Smartpillbottles} & 2013 & \catDxxWearable\catExxSmartHome & - & - & - & C & - & O,R & U & P \\ 
Whistle & \cite{Whistle} & \multicolumn{1}{l}{-} & \catDxxWearable\catExxSmartHome & $\checkmark$ & 99.95$^{1}$ & H,S & B,W & A,I & C & U & P \\ 
AirBoxLab & \cite{AirBoxLab} & 2013 & \catExxSmartHome & $\checkmark$ & - & H & W & A,I & C & U & P \\ 
BiKN & \cite{BiKN} & 2013 & \catExxSmartHome & $\checkmark$ & 129.99$^{1}$ & H,S & W & I & C & U & P \\ 
BrewBit & \cite{BrewBit} & 2012 & \catExxSmartHome & $\checkmark$ & 199.00$^{1}$ & H,S & W & A,I & O & U & P \\ 
Canary & \cite{Canary} & 2013 & \catExxSmartHome & - & - & H,S & W & A,I & C & U & P \\ 
Fliwer & \cite{Fliwer} & 2013 & \catExxSmartHome & $\checkmark$ & - & H,S & R,W & B & C & U & P \\ 
Good Night Lamp & \cite{GoodNightLamp} & 2013 & \catExxSmartHome & $\checkmark$ & 122.42-206.33$^{3}$ & H & W & B & C & U & P \\ 
Hintsights & \cite{Hintsights} & 2013 & \catExxSmartHome & - & - & H,S & - & - & O & - & S \\ 
iDoorCam & \cite{iDoorCam} & 2013 & \catExxSmartHome & $\checkmark$ & 164.95$^{1}$ & H,S & W & A,I & C & U & P \\ 
Iris & \cite{Iris} & 2013 & \catExxSmartHome & $\checkmark$ & 299.00$^{1}$ & H & W & B & C & U,S & P,S \\ 
Koubachi & \cite{Koubachi} & 2013 & \catExxSmartHome & $\checkmark$ & 122.41-273.72$^{3}$ & H & W & A,I & C & U & P \\ 
Lernstift & \cite{Lernstift} & 2013 & \catExxSmartHome & - & - & H & W & L & C & U & P \\ 
Lockitron & \cite{Lockitron} & 2009 & \catExxSmartHome & $\checkmark$ & 179.00$^{1}$ & H & B,W & A,I & C & U & P \\ 
Nest Thermostat & \cite{NestThermostat} & 2013 & \catExxSmartHome & $\checkmark$ & 249.00$^{1}$ & H,S & W & I,A & C & U & P \\ 
Netatmo & \cite{Netatmo} & 2013 & \catExxSmartHome & $\checkmark$ & 129.00-199.00$^{1}$ & H & - & A,I & C & U & P \\ 
Ninja Blocks & \cite{ninjablocks} & 2013 & \catExxSmartHome & $\checkmark$ & 199.00-250.00$^{1}$ & H,S & W,B,Z & A,B,I & C,O & U & P,S \\ 
OpenSprinkler & \cite{OpenSprinkler} & 2013 & \catExxSmartHome & $\checkmark$ & 114.15-206.31$^{3}$ & H,S & W & A,I & O & U & P \\ 
PetzillaConnect & \cite{PetzillaConnect} & 2013 & \catExxSmartHome & $\checkmark$ & 94.91-$^{3}$ & H,S & W & A,I & C & U & P \\ 
Philips Hue Connected Bulb & \cite{PhilipsHueConnectedBulb} & 2013 & \catExxSmartHome & $\checkmark$ & 199.95$^{1}$ & H,S & W & I & C & U & P \\ 
Pintofeed & \cite{Pintofeed} & 2013 & \catExxSmartHome & $\checkmark$ & 179.00$^{1}$ & H,S & W & A,I,W & C & U & P \\ 
Sensr.net & \cite{Sensr.net} & 2013 & \catExxSmartHome & $\checkmark$ & 94.91-$^{3}$ & H & W & B & C & U,S & P,S \\ 
SmartThings & \cite{SmartThingsHomeWatchSolution} & 2013 & \catExxSmartHome & $\checkmark$ & 199.00-299.00$^{1}$ & H,S & - & I,A & C & U & P \\ 
TADO & \cite{TADO} & 2013 & \catExxSmartHome & $\checkmark$ & - & H,S & W & I,A & C & U & P \\ 
Twine & \cite{Twine} & 2013 & \catExxSmartHome & $\checkmark$ & 124.95-199.95$^{1}$ & H & W & B & C & U & P \\ 
Ubi & \cite{Ubi} & 2013 & \catExxSmartHome & $\checkmark$ & 219.00$^{1}$ & H,S & W & A & O & U & P \\ 
WeMo Switch & \cite{WeMoSwitc} & 2013 & \catExxSmartHome & $\checkmark$ & 75.22-78.99$^{1}$ & H,S & W & A,I & C & U & P \\ 
WhereDial & \cite{WhereDial} & 2013 & \catExxSmartHome & $\checkmark$ & 162.77-179.20$^{4}$ & H & W & B & O & U & P \\


\hline
\end{tabular}
\vspace{-12pt}
\end{table*}

 \section{Trends and Lessons Learned}
 \label{sec:Lessons_Learned}
 In this section, we highlight and discuss some of the trends in the IoT solutions in the marketplace. The trends can be categorised \textit{1) based on domains, functionalities, and value}, and \textit{2) based on technology}.
 
  \subsubsection{Domains, Functionalities, and Value}
  \label{sec:TLL:Common_Domains}
 
  Most of the IoT solutions are narrowly focused on providing one functionality. However, we have seen that a number of generic platforms are being developed (e.g. Ninja Blocks \cite{ninjablocks}, SmartThings  \cite{SmartThingsHomeWatchSolution}, and Twine \cite{Twine}) to support applications in the domains of the smart home and the smart city. In general, more solutions are focused on the wearable and the smart home domains. One reason for this concentration is the market potential. These solution providers can earn a significant financial return for their solutions due to the larger consumer market. In addition, it has been revealed that the IoT solutions in the smart home and wearable domains are comparatively easy to develop and therefore low in price. Building smart enterprise, smart environment, and smart city solutions takes much effort and time due to the complexity and unique challenges in comparison to other domains. Some of the unique challenges are the sustainability of the hardware devices in harsh outdoor environments; the availability of energy sources for sensing, processing and communication in remote and outdoor locations; and the maintenance and repair of the hardware. These challenges justify the  low interest in these domain areas, specially on the part of  start-ups and small companies.

   \subsubsection{Technology: Hardware and Software Platforms}
   \label{sec:TLL:Hardware_and_Software_Platforms}
 
 According to the survey results presented in Table \ref{Tbl:Evaluation_of_Previous_Research_Efforts}, it is evident that most of the IoT solutions include both custom hardware and software. It is also to be noted that some of the solutions are not available for immediate purchase but are on the way to the market (e.g. pre-order). In terms of communication, WiFi and Bluetooth are the most commonly used protocols. Additionally, an increasing number of the IoT solutions support more than one platform (e.g. Android, iOS, browser-based, Windows, Linux, and Mac). Mostly, they are built around the Android and iOS platforms. Most of the solutions are protected under a commercial license and both software and hardware are closed-source. The majority of the IoT solutions are sold as units. Though solutions may have both software and hardware components, the price is mainly for the hardware and the accompanying software is free. The only exceptions are solutions that are completely based on the cloud, where they charge for subscription.

 In most of the wearable solutions, smartphones are used as an interface for human--system interaction. Smart wearable solutions generally have two or three components. Custom designed wearable devices are used to capture the context and sense the phenomena. Then, either processed or raw data is sent to a processing device, which is usually a smart phone (or a device with a similar computational capability). The smartphone then visualises and presents the outcome (e.g. alerts and notifications) to the users. One such example is Lumoback \cite{LUMOback}, which tracks posture and daily activities in real time. Lumoback collects data through a wearable waist belt and pushes the data directly to the smartphone. \textcolor{blue}{Human Computer Interaction (HCI) plays a significant roes in the success of IoT products and solutions. When combining different interaction mechanisms, IoT product designers will need  to select the right combination of methods based on number of different  factors such as data processing and communication capability, energy, hardware cost, target user knowledge, criticality of the product and so on. Commonly available options are gesture, voice, touch. Further, IoT products can use smart phones, tablets and wearable devices to enable user interactions.}
 
\begin{figure}[h!]
     \centering
     \includegraphics[scale=0.42]{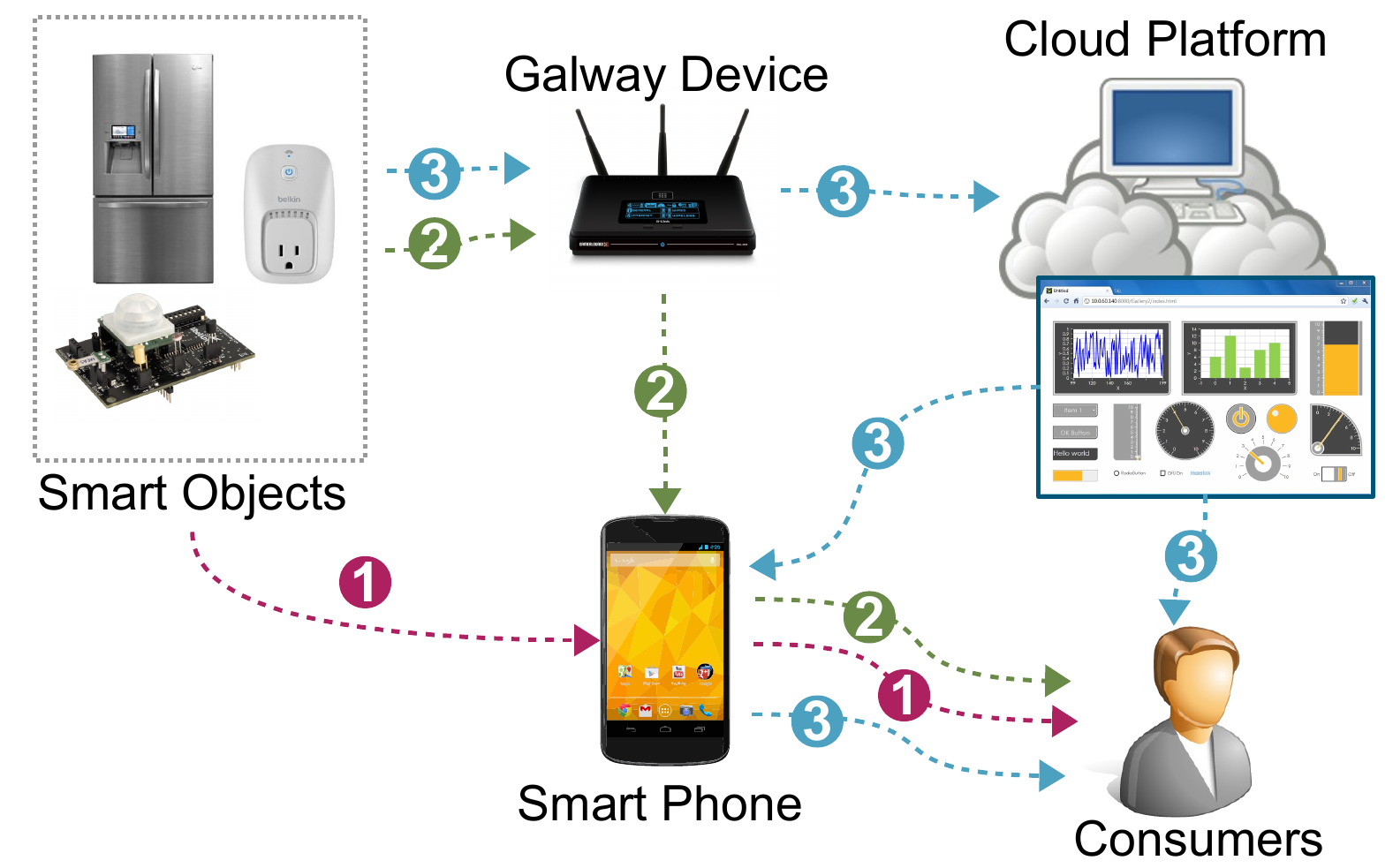}
    \vspace{-0.23cm}	
     \caption{Common Communication Patterns in IoT Applications. There are mainly three types of common patterns}
     \label{Figure:Common_Communication_Patterns}	
    \vspace{-0.32cm}	
    \end{figure}

 Alternatively, sensors may send data to custom gateway devices and then push to the cloud over GSM or WiFi. In such situations, cloud services push the outcome  to a mobile device to update the user on the real-time activities. For example, Mimobaby \cite{RestDevice} is a baby movement monitoring wearable solution. Mimobaby collects data from sensors attached to the baby's clothes. Then, it transfers the data to a nearby custom gateway which uses home WiFi connectivity to push the data to the cloud. Then, the cloud services alert the parents' smartphone  in real-time. Figure \ref{Figure:Common_Communication_Patterns} illustrates some of the most common communication patterns used in the IoT solutions. Data collected by the IoT solutions may be sent to the cloud for further processing, historical archiving, or pattern recognition. Mobile devices allow users to immediately take action or perform actuation tasks. In such circumstances, the communication between the hardware and the mobile devices is performed using short distance communication protocols, such as Bluetooth, and long range communication tasks are performed via WiFi or GSM. 
   
   It is also evident that  cloud IoT platforms are trying to build their own ecosystems by facilitating and supporting third party extensions (also called plugins) development and distribution through \textit{app store}. We have repeatedly seen such trends in both PC market and smartphone markets. IoT platform developers are increasingly support non-technical people to build IoT solutions by providing easy ways to assemble the components without programming knowledge \cite{WaspMote, Z1069}.

  \section{Open Research Challenges}
  \label{sec:Ope_Research_Challenges}
 \setcounter{subsubsection}{0}
 
 In this section, our objective is to discuss some of the major challenges that need to be addressed in order to  build the IoT. These challenges are yet to be  addressed, by either academia or industry, in a comprehensive manner. The solutions for these issues need to be come from technological, social, legal, financial, and business backgrounds in order to receive wide acceptance by the IoT community.

 \subsubsection{Modularity and Layered Interoperability}
 \label{sec:ORC:Modularity}
 Modularity is a key to success in the IoT paradigm. The notion of modularity goes hand in hand with interoperability. We can interpret modularity in different ways. First, the hardware / physical layer in the IoT needs to support modularity. This means, ideally, consumers should be able to build a smart object (or an Internet connected object) by putting different modules (e.g. sensors or actuators) produced by different manufacturing companies together without getting restricted to one vendor. Such modularity reduces the entry barriers to  the IoT marketplace. Further, interoperability will increase the level of  creativity and will reduce costs due to competition. Modularity also allows organisations to focus on one component of the IoT architecture and become experts on that, rather than having to build end to  end solutions, something which leads to re-inventing the wheel. Furthermore, modularity provides more choices and options to the consumers as to which modules to use and when, based on factors such as reliability.
 
 Modularity is vital in software / cloud services layer as well. Especially in sensing as a service model \cite{ZMP008}, something which provides an economical business model for the IoT, users should have the right and flexibility to choose and use cloud services either in a standalone fashion or as a composition of multiple services \cite{Z1062, Z1060} based on their own priorities and preferences \cite{ZMP009}. Modularity needs to be governed by rigorous standardisation processes. Semantics technologies can also be used to improve the interoperability through knowledge reuse and knowledge mapping. In addition, interoperability can be achieved through mediators and adapters. At the hardware level, modularity has been introduced to some extent by platforms such as Arduino \cite{P411} and Microsoft's .NET Gadgeteer \cite{Z1055}. However, cross platform compatibility is not yet supported. In recent years, we have seen many different IoT companies building cross layer partnerships with each other to ensure comparability and interoperability. For example, sensing hardware platform designers are partnering with cloud IoT solution providers. However, standardisations and collaboration with competitors is rarely seen within layers (e.g. among different hardware vendors). The concept of an app store for IoT solutions is currently supported by HomeOS \cite{Z1058} and ThingWorx \cite{thingworx}. They have started to support modularity by allowing third parties to develop extensions to their IoT middleware platforms. The integration of multiple cloud platform service providers will enable more data sharing and value creation \cite{Z1063}.

  \subsubsection{Unified Multi-Protocol Communication Support}
  \label{sec:ORC:Unified_Multi-Protocol_Support}
  Designing protocols and systems  for wireless industrial communications will have a significant impact on the successful adoption of the IoT \cite{Z1068}. IoT solutions use different types of communication protocols, mostly through wireless channels \cite{Z1014}.  WiFi, Bluetooth, 3G, Zigbee, and z-wave are some of them. Even though they seem few, incompatibility makes developing the IoT  applications more challenging. Each protocol has its own advantages and disadvantages \cite{Z1067}. Comparisons of these protocols are presented in \cite{Z1014, Z1067}. Some protocols are efficient in long distance communication and others are efficient in short distance communication. It is important to address the challenge of developing a high-level framework that handles the difference of protocols behind the scenes without bothering the developers or consumers. Therefore, an ideal framework should allow the developers to focus on data communications at a high level (e.g. what to send  and when) rather than dealing with low-level communication protocol details (e.g. which protocol to use when, and implementation-level differences). Such a high-level framework will increase the efficiency and effectiveness of the IoT solutions and will also save a significant amount of development time. Intelligent context-aware  capabilities need to be integrated into the IoT solutions so the communication tasks and related decisions will be made and handled based on the capabilities and the energy availability of the ICOs at a given situation. The importance of standardising a protocol stack for the IoT is highlighted and discussed in \cite{Z1041}.
 
  \subsubsection{Sustainable Business Models}
  \label{sec:ORC:Business_Models} 
  
 A sustainable business model is essential for building a sustainable IoT paradigm. Most of the IoT solutions we have reviewed are narrowly focused on addressing one problem. They have missed the bigger picture of the IoT and earnings potential. Sharing data in open markets can add more value to the IoT solutions. One such business model is presented in \cite{ZMP008} in detail. Preliminary work towards building such a model is currently conducted by the project HAT (Hub of All Things) \cite{Z1056}. HAT is a platform for a multi-sided market powered by the IoT which expects to create opportunities for new economic \& business models. It aims to create a market platform for the home based on the data generated by individuals' consumption, behaviour, and interactions. Such data exchange, in secure and privacy preserved manner, would generate additional value that may help to maintain the IoT infrastructure in the long-term. 
 
 For example, one institution (primary) may  deploy and maintain sensors in public infrastructures such as roads and bridges with the intention of monitoring their structural health and the public safety. Other institutions (secondary) who are interested in such data can purchase them through an auction-like marketplace.  Secondary institutions may have different intentions, such as local weather monitoring, environmental pollution monitoring, and  local traffic condition monitoring. The financial support offered by secondary institutions motivates the primary institute to deploy and maintain sensors over the long term.   In this way, primary and secondary institutions will benefit by the transaction's creating a sustainable economic model. The details of such markets are discussed from the technology perspective in \cite{ZMP008} and from the business perspective in \cite{Z1056}.

   \subsubsection{Ownership, Privacy and Security}
   \label{sec:ORC:Ownership, Privacy and Security} 
 One of the biggest technical, social, and legal challenges is protecting privacy and creating a secure environment for the IoT. Unfortunately, these have been the challenges least addressed. Due to the limited adoption of IoT, not many security and privacy challenges have been identified. We can expect more challenges to be identified over the coming years due to the growing adoption of IoT solutions. The security issues have two aspects. One aspect is data security. The other aspect is the security of the IoT solutions  (e.g. security related to sensing communication, iterations, authentication, and actuation). In the fully automated and integrated IoT paradigm, security breaches can be life threatening and can have devastating economic and social impact. Especially the new business models that we briefly discussed in the above section may create additional challenges regarding data ownership and privacy. 
 
 As we discussed in \cite{ZMP008}, anonymisation is a critical process in the IoT data flow. The data collected by households always needs to be anonymised in such a way that no one will be able to trace it back to its exact origin. Data may identified and grouped broadly  into certain geographical regions, but not for individuals or households. Another aspect of this challenge is ownership transfer. Technology should be intelligent enough to identify its current owner and follow their commands and preferences. The details of such ownership transfers are discussed in \cite{ZMP008}. In addition to the technology-based security and privacy solutions, legal terms need to be developed in order to protect the consumers and the data they own.

 
 \section{Concluding Remarks}
 \label{sec:Conclusions}
 
 This paper presented a survey of the IoT solutions in the emerging marketplace. We classified the solutions in the market broadly into five categories: smart wearable, smart home, smart city, smart environment, and smart enterprise. Under each category, we discussed and summarised the functionalities provided by each solution. We also examined the contribution of each solution towards improving the efficiency and effectiveness of consumers' lifestyle as well as  of society in general. It is important to highlight the proliferation of wearable solutions in the market. Despite the long existence of wearable computing, those products did not reach the consumer market until recently. It is clear that more and more wearable solutions will make their way into the IoT marketplace over the coming years. Further, we can see a significant investment and focus on indoor smart home and office domains, in comparison to environmental monitoring  solutions. 
 
 Moreover, we also see a substantial amount of investment made in research and development towards supply chain management. These solutions are aimed at large scale industry players who are looking for novel methods to optimise their supply chain processes, especially through real-time data collecting, reasoning, and monitoring.  Until household consumers adopt IoT solutions, the majority of the value creation is expected to occur with large scale industries. Finally, we discussed the lessons learned and listed some of the major research challenges and opportunities. We believe further research that addresses these open challenges will help to develop more interesting IoT solutions and strengthen the existing solutions in this area in both the industrial and the academic sectors.
 
%


%
 
\ifCLASSOPTIONcaptionsoff
  \newpage
\fi




\bibliography{Bibliography}
\bibliographystyle{IEEEtran}

%
%

%

%


%

%
%
\vspace{-20pt}
\begin{IEEEbiography}[{\includegraphics[width=1in,height=1.25in,clip,keepaspectratio]{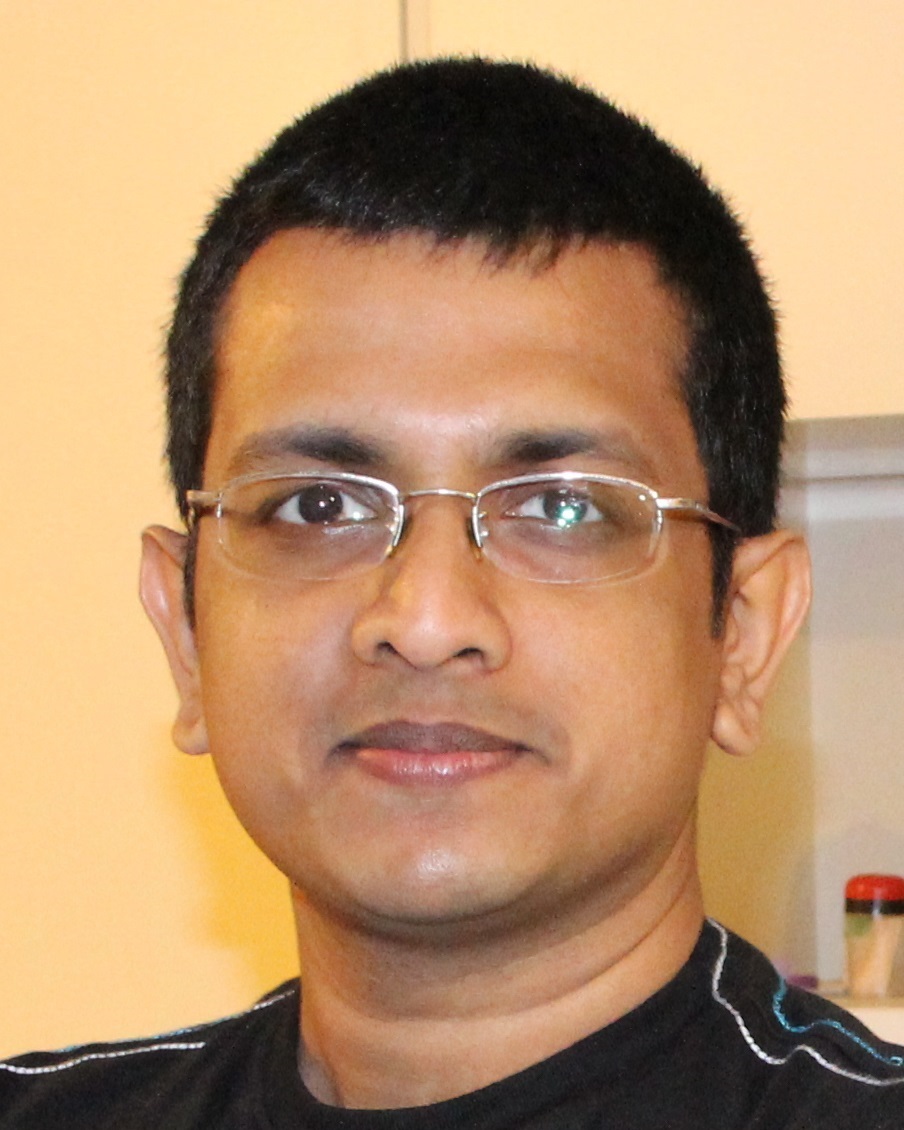}}]{Charith Perera}
received his BSc (Hons) in Computer Science in 2009 from Staffordshire University, Stoke-on-Trent, United Kingdom and MBA in Business Administration in 2012 from University of Wales, Cardiff, United Kingdom and PhD in Computer Science at The Australian National University, Canberra, Australia. He is also worked at Information Engineering Laboratory, ICT Centre, CSIRO and involved in OpenIoT Project (FP7-ICT-2011.1.3) which is co-funded by the European Commission under seventh framework program. He has also contributed into several projects including EPSRC funded HAT project (EP/K039911/1)  His research interests include Internet of Things, Smart Cities, Mobile and Pervasive Computing, Context-awareness, Ubiquitous Computing. He is a member of both IEEE and ACM.
\end{IEEEbiography}

\vspace{-20pt}

\begin{IEEEbiography}[{\includegraphics[width=1in,height=1.25in,clip,keepaspectratio]{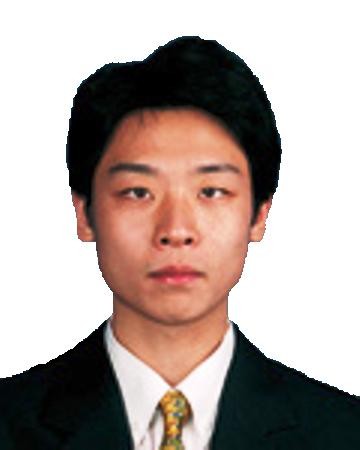}}]{Chi Harold Liu}
is a Full Professor at the School of Software, Beijing Institute of Technology, China. He is also the Deputy Director of IBM Mainframe Excellence Center (Beijing), Director of IBM Big Data Technology Center, and Director of National Laboratory of Data Intelligence for China Light Industry. He holds a Ph.D. degree from Imperial College, UK, and a B.Eng. degree from Tsinghua University, China. His current research interests include the Internet-of-Things (IoT), big data analytics, mobile computing, and wireless ad hoc, sensor, and mesh networks.  He served as the consultant to Bain \& Company, and KPMG, USA, and the peer reviewer for Qatar National Research Foundation, and National Science Foundation, China. He is a member of IEEE and ACM.

\end{IEEEbiography}

\vspace{-20pt}

\begin{IEEEbiography}[{\includegraphics[width=1in,height=1.25in,clip,keepaspectratio]{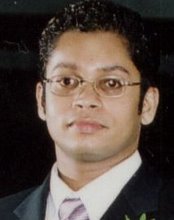}}]{Srimal Jayawardena}
received his BSc (Hons) in Electrical Engineering from the University of Peradeniya and his Bachelors in IT from the University of Colombo School of Computing, both with first class honours in 2004. He also obtained a Masters in Business Administration from the University of Moratuwa in 2009. He holds a PhD in Computer Science from The Australian National University, Canberra. He is currently working as post-doctoral research fellow at Computer Vision Laboratory (CI2CV),in CSIRO. His research interests include augmented reality, object recognition for the Internet of Things, computer vision, human computer interaction, and machine learning. He is a member of the Institute of Electrical and Electronics Engineers (IEEE).
\end{IEEEbiography}









\end{document}